\documentclass[pdflatex,sn-mathphys-num]{sn-jnl}

\usepackage{multirow}%
\usepackage{amsmath,amssymb,amsfonts}%
\usepackage{mathrsfs}%
\usepackage[title]{appendix}%
\usepackage{xcolor}%
\usepackage{textcomp}%
\usepackage{manyfoot}%
\usepackage{booktabs}%
\usepackage{algpseudocode}%
\usepackage{listings}%

\usepackage{algorithm}
\usepackage{amsfonts,amssymb}
\usepackage{subfig}
\usepackage{float}
\usepackage{graphicx}

\begin{document}

\title[Article Title]{CIDER: Counterfactual-Invariant Diffusion-based GNN Explainer for Causal Subgraph Inference}


\author[1]{\fnm{Qibin} \sur{Zhang}}\email{i@zhangqibin.com}

\author[2]{\fnm{Chengshang} \sur{LYU}}\email{chengshang.lyu@cityu.edu.hk}

\author[2]{\fnm{Lingxi} \sur{Chen}}\email{lingxi.chen@cityu.edu.hk}

\author[3]{\fnm{Qiqi} \sur{Jin}}\email{jinqiqi@sibcb.ac.cn}

\author*[1, 3, 4]{\fnm{Luonan} \sur{Chen}}\email{lnchen@sibcb.ac.cn}

\affil[1]{\orgdiv{Key Laboratory of Systems Health Science of Zhejiang Province, School of Life Science}, \orgname{Hangzhou Institute for Advanced Study, University of Chinese Academy of Sciences}, \orgaddress{\city{Hangzhou}, \postcode{310024}, \country{China}}}

\affil[2]{\orgdiv{Department of Biomedical Sciences}, \orgname{City University of Hong Kong}, \orgaddress{\city{Hong Kong}, \country{China}}}

\affil[3]{\orgdiv{Key Laboratory of Systems Biology, Shanghai Institute of Biochemistry and Cell Biology}, \orgname{Center for Excellence in Molecular Cell Science, Chinese Academy of Sciences}, \orgaddress{\city{Shanghai}, \postcode{200031}, \country{China}}}

\affil[4]{\orgname{Guangdong Institute of Intelligence Science and Technology}, \orgaddress{\city{Hengqin, Zhuhai}, \postcode{519031}, \country{China}}}


\abstract{Inferring causal links or subgraphs corresponding to a specific phenotype or label based solely on measured data is an important yet challenging task, which is also different from inferring causal nodes. While Graph Neural Network (GNN) Explainers have shown potential in subgraph identification, existing methods with GNN often offer associative rather than causal insights. This lack of transparency and explainability hinders our understanding of their results and also underlying mechanisms. To address this issue, we propose a novel method of causal link/subgraph inference, called CIDER: Counterfactual-Invariant Diffusion-based GNN ExplaineR, by implementing both counterfactual and diffusion implementations. In other words, it is a model-agnostic and task-agnostic framework for generating causal explanations based on a counterfactual-invariant and diffusion process, which provides not only causal subgraphs due to counterfactual implementation but reliable causal links due to the diffusion process. Specifically, CIDER is first formulated as an inference task that generatively provides the two distributions of one causal subgraph and another spurious subgraph. Then, to enhance the reliability, we further model the CIDER framework as a diffusion process. Thus, using the causal subgraph distribution, we can explicitly quantify the contribution of each subgraph to a phenotype/label in a counterfactual manner, representing each subgraph's causal strength. From a causality perspective, CIDER is an interventional causal method, different from traditional association studies or observational causal approaches, and can also reduce the effects of unobserved confounders. We evaluate CIDER on both synthetic and real-world datasets, which all demonstrate the superiority of CIDER over state-of-the-art methods.}

\keywords{Causal Inference, Bioinformatics, Graph Neural Network}



\maketitle

\section{Introduction}
How to identify causal links or subgraph with respect to the samples' labels based only on measured data, attracts a great amount of attention in various communities, and such a problem is also different from inferring causal nodes. Graph Neural Networks(GNNs) have emerged as a powerful tool in handling and learning graph-structured information for such a task. As a variant of neural networks, GNNs have been crafted to function on graph-structured data, including molecular compositions, social connections, and knowledge graphs.

Recently, Graph Neural Networks Explainer (GNNs Explainer) has shown potential in identifying subgraphs that associate with a specific GNN's output, but it provides only explanations at a model level. In other words, it would explain the association between subgraphs and the specific task GNN, rather than answering the important question \textit{"Which part of the graph results in such a label?"}. 

For example, \cite{ying2019gnnexplainer} and \cite{luo2020parameterized} estimate the contribution of each subgraph to the trained GNN's output, which can help to understand the GNN's mechanism but may fail in finding the causal subgraph that really affects phenomena or labels. On the other hand, \cite{lin2021generative} and \cite{lin2022orphicx} aim to provide causal information by using causal theory, but the former considers each edge's contribution to the trained GNN's output by using a greedy algorithm to search for the most important subgraph without the consideration of the dependence between edges. The latter first maps the graph to a hidden space to compute the contribution of every causal factor to the trained GNN's output, and then uses the causal factors to reconstruct the graph's structure. However, the dependence between those factors is ignored, and thus, they may result in the reconstructed subgraph with no causal links. Therefore, in a low-sparsity situation, they have poor results.

\begin{figure*}[t]
    \centering
    \subfloat[Illustration of CIDER.]{
        \includegraphics[width=0.9\linewidth]{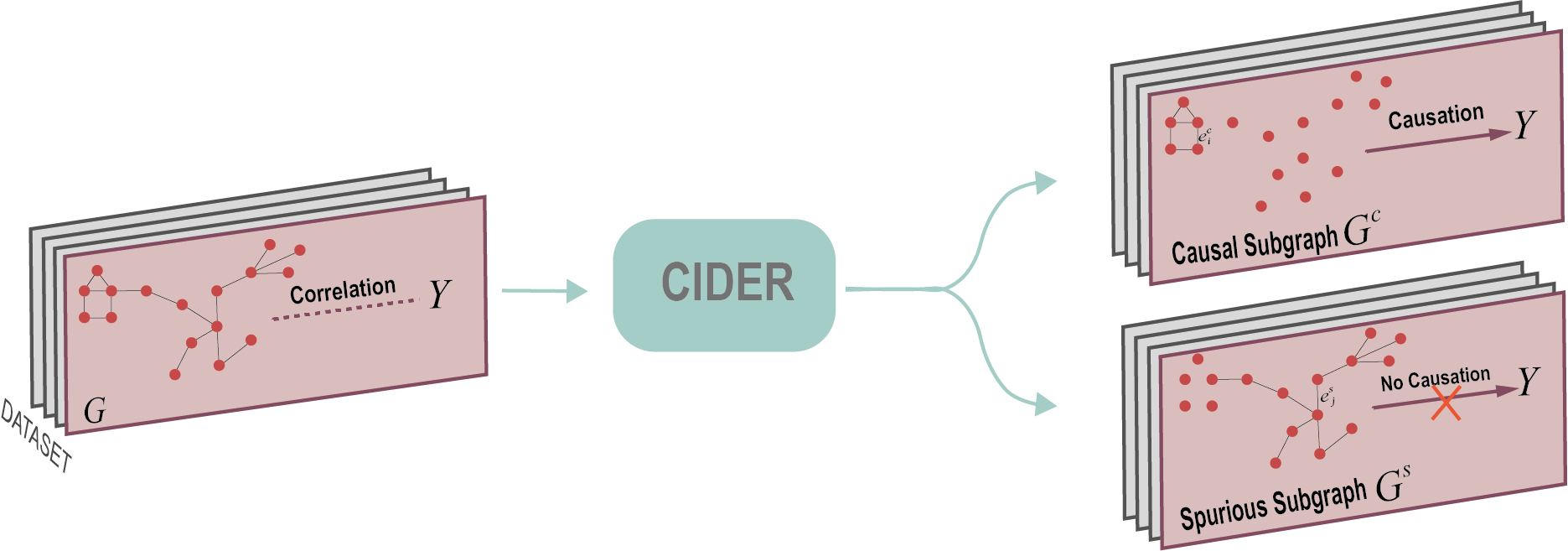}
        \label{fig.example}
    }
    \quad
    \subfloat[Framework of CIDER.]{
        \includegraphics[width=1\linewidth]{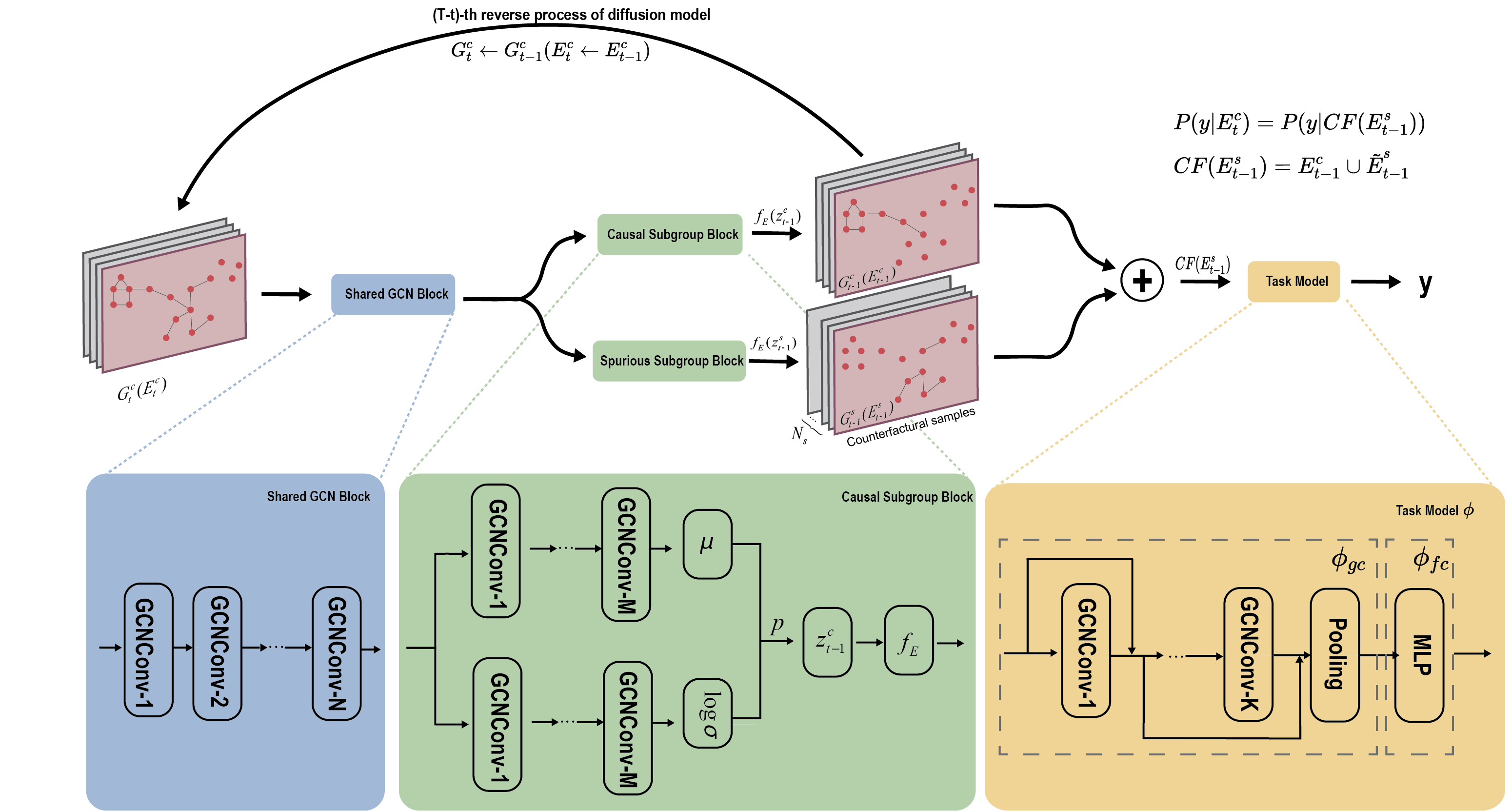}
        \label{fig.total}
    }
    \caption{Illustration and framework of CIDER. (a) Conceptual illustration of CIDER; (b) Framework of CIDER. For a graph with a label $Y$ is given, CIDER gives the causal subgraph, which causally affects the label, and the spurious subgraph, which does not affect the label.}
    
\end{figure*}

To provide a causal subgraph from data, we propose a novel framework called CIDER: Counterfactual-Invariant Diffusion-based GNN ExplaineR, which can answer both \textit{"Which part of the graph causally affects a sample's label?"} and \textit{"Which part of the graph results in the trained GNN's output?"}. CIDER is an intervention and Counterfactual-Invariant method that trains a diffusion-based model to divide the whole graph or network into one causal subgraph and one spurious subgraph. That is, it directly predicts the distributions of causal and spurious subgraphs, enabling the estimation of causal strength, i.e., each subgraph's contribution to the graph's label. Figure (\ref{fig.example}) shows one example of CIDER for every graph with a label, where CIDER gives the causal subgraph and spurious subgraph. Note that we can use the measured data $X$ to construct a graph (e.g., by correlation analysis) if the graph is not explicitly given. Our major contributions can be highlighted below.

We propose a novel framework called CIDER, which provides a causal explanation from a subgraph to a label or phenotype by training a GNN based only on the measured high-dimensional data. By implementing intervention and Counterfactual-Invariant to train a diffusion model, CIDER directly constructs the two distributions of one causal subgraph and another spurious subgraph, i.e. by dividing the whole graph into these two subgraphs, thus quantifying the contribution of each subgraph to the graph's label. Specifically, to achieve this, we first model the problem as a variant inference task that predicts the two subgraph distributions, as shown in Figure \ref{fig.example}. Then, we use a VGAE-based framework with two channels to predict the distributions: one for the causal subgraph and the other for the spurious subgraph. The two distributions are further used to reconstruct the original graph's structure, and the contribution of each subgraph to the label is computed by doing sample counterfactual implementation in the spurious subgraph. Note that CIDER is an interventional causality method, different from traditional association studies or observational causality methods, and in particular, is able to reduce the influence of unobserved confounders, which is actually a long-standing problem in the field.  

CIDER adopts a diffusion process to further enhance the inference capability, where denoising and distilling the spurious subgraph are performed at each diffusion step. After the T steps, each step's causal subgraph distribution and noise distribution are obtained. We use the noise distribution's significant attribute and counterfactual invariant to make CIDER trainable, even though the noise distribution is not required in the traditional diffusion model's training process. This results in a sparser and more reliable causal subgraph distribution.

CIDER provides not only a causal subgraph due to counterfactual implementation but also its causal links owing to the diffusion process. Experimental results on both synthetic and real-world datasets, including single-cell RNA-seq datasets, demonstrate such advantages, which shows that our CIDER outperforms state-of-the-art methods. 

\section{Results}
This section presents the theoretical and computational results of CIDER with a comprehensive set of experiments evaluating CIDER's performance. We begin outlining the result of the CIDER theory. Subsequently, we compare CIDER against several state-of-the-art methods on both benchmarks. Finally, we propose a framework to use CIDER to analyze single-cell RNA-seq data, illustrating its efficacy and applicability in real-world scenarios. The code and data utilized in this work are readily accessible in the supplementary material. 

\subsection{Theoretical results of CIDER to ensure interventional causality}
We first show the theoretical result of CIDER, which is able to infer the causal subgraph (or a set of edges) against an output label or phenotype of a sample based only on data in the sense of interventional causality. The conceptual illustration is given in Figure 1. 

We define a random graph as $G = (X, V, E)$ where $|E|$ and $|V|$ are numbers of edges and nodes, respectively. Each node has a $d$-dimensional feature vector, and $X$ is a $|V| \times d$ matrix (the measured data of a sample or graph). We aim to derive the conditional distribution of the edges, $P(E|X,V)$ from the data, which is represented as $P(E)$ for simplicity. As described in detail in Methods, by examining the relationship between the distribution of $E^c_0$ that is the resultant causal subgraph $E^c$, and $E^c_T$ that is the (initial) original graph $E$ against label ($y$), we can obtain Eq.(\ref{eq.ect}). Then, for the $(T-t)$-th backward step (i.e., from T to 0) of the inference process regarding the distribution of $E^c_t$, we derive the main theoretical result Eq.(\ref{eq.couterfactual}): interventional causality from the causal subgraph $E^c$ to the label $y$ of the measured samples, in the sense of counterfactual intervention. 
\begin{eqnarray}
    \label{eq.ect}
    P(E^c_T)&=&P(E^c_0)\prod_{t=0}^{T-1} P(E^s_t|E^c_t) \\
    \label{eq.couterfactual}
    P(y|CF(E^s_t)) &=& P(y|E^c_{t+1})
\end{eqnarray}
where  $CF(E^s_t) = E^c_t \cup \tilde{E}^s_t$, and $\tilde{E}^s_t$ denotes a counterfactual sample derived from $P(E^s_t|E^c_t)$. This result indicates that the converged $E^c$ with the step or iteration $t$ (i.e., $t=0,1,...,T-1$) is the causal subgraph for label $y$, where $E^s$ is the remaining (resultant) non-causal subgraph among the original graph $E$. In order to simultaneously infer $P(E^c_t)$ and $P(E^s_t|E^c_t)$, we utilize a two-channel VGAE as shown in Figure \ref{fig.total}. Figure \ref{fig.total} depicts the causal inference block, which fits $P(E^c_{t-1}|E^c_{t})$. We can have $N, M$ and $K$ layers of GCNconv for Sharing GCN Block, Causal Infer Block and Task Model, where $N, M, K$ are dataset dependent. In general, a model with more layers could have better performance. However, we use only one GCN layer in the Shared GCN Block and Infer Block for simplicity and ease of interpretation. Note that the Spurious Infer Block has the same structure as the Causal Infer Block. 

\subsection{Computational results on benchmarks}
We mainly follow the experimental settings of the prior work \cite{ying2019gnnexplainer, lin2022orphicx} with some differences due to our goal of explaining both the GNN output and the underlying causal factors. As previously mentioned, we achieve the explainer task at a graph-level and thus evaluate our model on graph classification tasks. We use one synthetic dataset BA-2motif\cite{luo2020parameterized} and two real-world datasets, MUTAG and NCI1, both widely used in graph classification, and summarize their statistics in Appendix.

BA2-motif: This synthetic dataset, constructed by \cite{luo2020parameterized}, utilizes BA graphs as its base. These base graphs are generated through a Markov process, with half subsequently attached to "house" motifs and the remaining half attached to five-node cycle motifs. Each graph is then assigned to one of two classes based on the attached motif type.

\subsubsection{Baselines and experimental setup}
We compare our model with the following baselines: GNNEplainer\cite{ying2019gnnexplainer}, OrphicX\cite{lin2022orphicx}, GEM\cite{lin2021generative}, PGExplainer\cite{luo2020parameterized}. We evaluate our model not only in explaining the task GNN model, but also in causally explaining the phenomenon itself. Hence, we conduct two parts of experiments: one is the traditional GNN explaining experiments where we use the origin task GNN's output as a criterion (the fewer edges used to predict, the closer to the origin output, the more powerful) without the consideration of the labels; the other one uses the ground truth of the labels as a criterion (the more edges used to predict the ground truth, the better).

\subsubsection{Results on synthetic datasets}
In this section, we show CIDER's performance in the BA-2motif dataset. The BA-2Motif Dataset, featured in the \cite{luo2020parameterized}, serves as a synthetic dataset crafted for the assessment of explainability algorithms. This dataset comprises 1,000 randomly generated Barabási-Albert (BA) graphs, a selection designed to mirror the complexity and variability inherent in graph-based data structures. To introduce further complexity and to facilitate the evaluation of algorithmic explainability, these graphs are augmented with one of two distinct motifs: a HouseMotif or a five-node CycleMotif.

The presence of these motifs divides the dataset into two classifications. Graphs adorned with a HouseMotif fall into one category, while those featuring a five-node CycleMotif are allocated to the other. This bifurcation is central to the dataset's utility in testing the efficacy of explainability algorithms, as it presents a clear, binary classification challenge based on the type of motif attached to the underlying BA graph. 

In the context of the synthetic BA-2Motif Dataset, CIDER demonstrates exceptional performance, achieving nearly 100\% accuracy in classifying graphs based on the attached motifs. This remarkable level of precision underscores CIDER's effectiveness and power in addressing complex pattern recognition challenges within graph data. 

\subsubsection{Results on real-world datasets}
In this section, we aim to answer the question of which parts of the graph causally contribute to the label or phenotype. As our goal is the underlying phenomena and causality, we evaluate our model based on its ability to predict the ground truth of the labels by training a GNN from the observed data. To this end, we modify the loss function $L_1$ to $\sum_{i=1}^{N_c}\sum_{j=1}^{N_s} \left| \phi_{fc}(\phi_{gc}(X, CF(E^s)) - y \right|$. The results of our model and baselines on explaining phenomena tasks are shown in Table \ref{tab.results1}. We observe that our model outperforms all the baselines on all datasets. This indicates that our model can learn the causality from the subgraph to underlying phenomena. Consequently, our model can use fewer edges to predict the ground truth.

\begin{table}[ht]
    \caption{Accuracy in causally explaining phenomena on real-world datasets(\%)}
    \label{tab.results1}
    \centering
    \begin{tabular}{l||llll|llll}
      \toprule
      &\multicolumn{4}{c|}{\textbf{MUTAG}} & \multicolumn{4}{c}{\textbf{NCI1}}\\
      Sparsity &0.1 & 0.2 & 0.3 & 0.4 &0.1 & 0.2 & 0.3 & 0.4\\
      \midrule
      \textit{GNNEplainer} & 0.492  & 0.532  &  0.58 & 0.620 & 0.502  & 0.544  &  0.532 & 0.528   \\
      \textit{PGExplainer}  & 0.500  & \underline{0.554}  &  0.578 & \underline{0.648} & 0.512  & 0.510  &  0.502 & 0.514  \\
      \textit{GEM} & 0.57  & 0.580  &  0.603 &  0.624 & 0.537  & 0.562  &  0.560 & 0.557  \\
      \textit{OrphicX} & 0.442  & 0.479  &  0.534 & 0.582 & 0.506  & 0.520  &  0.491 & 0.501   \\
      \textit{CIDER} & \textbf{0.640} & \textbf{0.640} & \textbf{0.672} & \textbf{0.674} & \textbf{0.676} & \textbf{0.680} & \textbf{0.692} & \textbf{0.688} \\
      \bottomrule
    \end{tabular}
  \end{table}

\subsection{Real-world applications}
  In this section, we demonstrate the application of CIDER in analyzing COVID-19 single-cell RNA-seq (scRNA-seq) data and AML bulk RNA-seq data, highlighting its potential and adaptable utility in real-world scenarios. The primary advantage of utilizing CIDER for RNA-seq data analysis lies in its capacity to simultaneously elucidate both causal genes and CellType network. This capability significantly facilitates the identification of underlying biological mechanisms at a molecular level.
  
\subsubsection{COVID-19 dataset}
We applied CIDER to a scRNA-seq dataset from COVID-19\cite{ahern2022blood}, which includes gene expression profiles from 8,361,000 single cells obtained from the blood of 124 individuals in three different states: COVID-19, influenza, and normal. These single cells have been annotated into 19 distinct cell types in the original study.

\subsubsection{Metric evaluation on the COVID dataset}
Initially, we employed a Graph Neural Network (GNN) for the COVID-19 gene network, consisting of 370 nodes (genes) and 498 edges, achieving a graph classification accuracy of 0.85. Following the application of CIDER, the network was significantly streamlined to merely 30 edges, while maintaining a classification accuracy of 0.82 with the same GNN. This result indicates that CIDER preserved approximately 96\% of the performance in deducing gene-gene causal subgraphs, while substantially reducing the network's complexity for inference tasks. Moreover, CIDER managed to retain 91\% of the performance in identifying cell type-cell type causal subgraphs, demonstrating its efficiency in network sparsification without significant loss of accuracy.

\subsubsection{Biological analysis for COVID-19 dataset}
To further elucidate the effectiveness of our CIDER method in the real-world data, we conducted additional validation using the COVID-19 RNA-seq dataset from blood\cite{ahern2022blood}. Our CIDER method allowed us to refine the top $1\%$ of highly variable genes (375 genes) in the dataset, from which we selected the top 30 genes ($<0.08\%$) with the highest causal association scores as key genes for biological analysis using tools like the Database for Annotation, Visualization and Integrated Discovery (DAVID)\cite{huang2009david} and STRING database\cite{szklarczyk2023string}.

\begin{figure*}[ht]
    \centering
    \subfloat[Key Genes' Network Generated by StringDB]{\includegraphics[width=.5\linewidth]{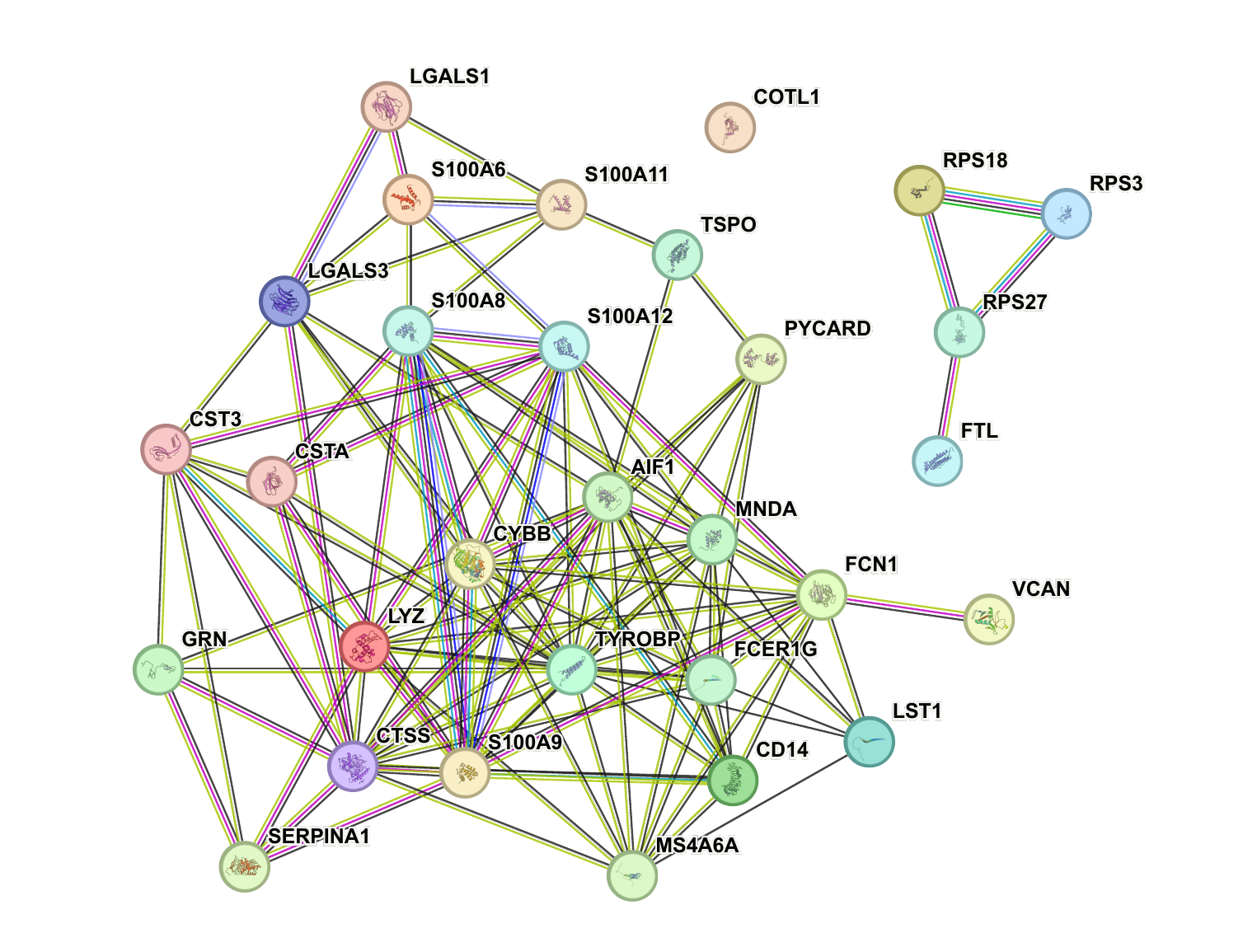} \label{fig.6A}}
    \subfloat[KEGG Pathway Enrichment Circle Graph]{\includegraphics[width=.5\linewidth]{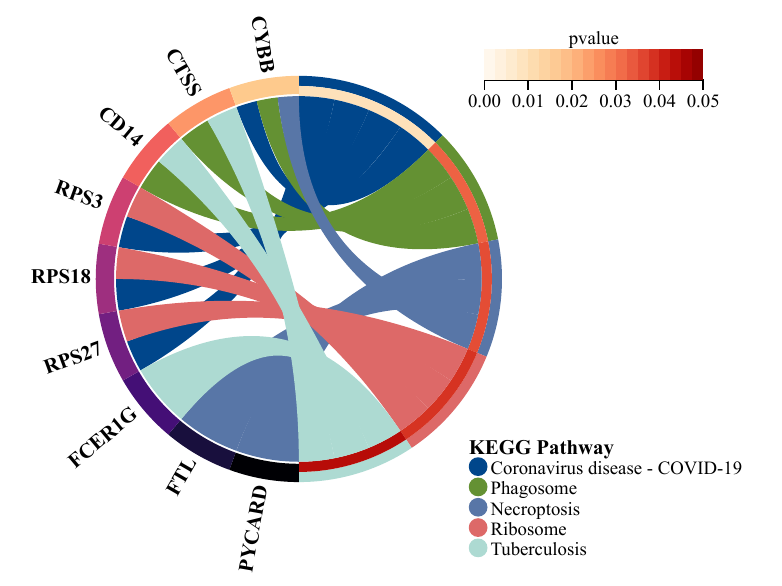}  \label{fig.6B}}
    \hspace{2pt}
    \subfloat[Gene Ontology Enrichment Circle Graph]{\includegraphics[width=.5\linewidth]{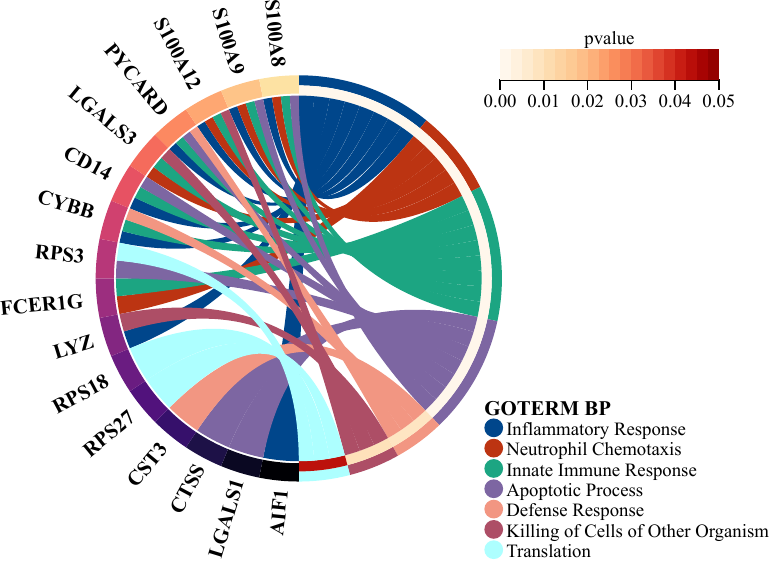} \label{fig.6C}}
    \subfloat[Uniprot Keywords Enrichment Circle Graph]{\includegraphics[width=.5\linewidth]{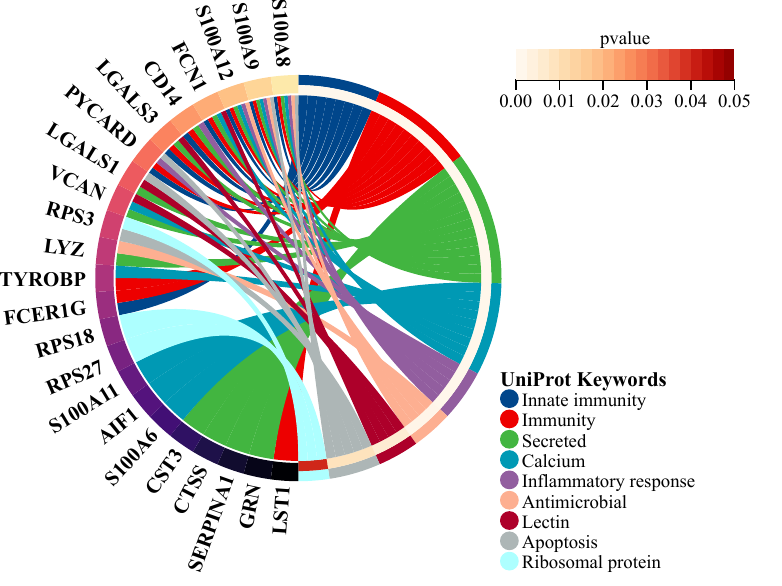} \label{fig.6D}}
    \caption{COVID19 scRNA-seq analyses by CIDER. (a) Three subgraphs generated by STRING illustrate the division of 30 key genes into three distinct clusters, revealing potential unique functions and interactions within the molecular network related to COVID-19. (b) KEGG Pathways related to ribosome, apoptosis, tuberculosis, and COVID-19 were enriched, suggesting a direct connection between the key genes and the disease mechanisms. (c and d) Most genes were enriched in immunity-related keywords, supporting the effectiveness of the CIDER method in identifying COVID-19-related genes. It further emphasized the role of the key genes in the inflammatory conditions of COVID-19 and their potential prognostic value for severe cases.}
\end{figure*}

In the STRING network (Figure \ref{fig.6A}) with a score threshold $>0.4$, it is apparent that these 30 genes are partitioned into 3 independent subnetworks. One large subnetwork comprising genes such as CYBB and the S100 protein family, a smaller subnetwork formed by the RPS family of genes alongside FTL, and the standalone gene COTL1. These three separated subnetworks potentially indicate three or more distinct network functions, which will be a focal point in our further research work in combination with other analytical tools.

The isolated COTL1 may have a unique, yet unknown impact. Previous studies have reported the role of COTL1 in the immune response\cite{wang2022coactosin,kim2014coactosin,xia2018coactosin} and were also mentioned involved in the SuperPath's Innate Immune System\cite{belinky2015pathcards}, while the COTL1 protein in the Airway mucus of COVID-19 patients was shown to be downregulated\cite{zhang2022proteomic}.

KEGG Pathway\cite{kanehisa2000kegg} enrichment results (Figure \ref{fig.6B}) revealed routes including ribosomes, apoptosis, tuberculosis, and COVID-19, hinting at a possible direct linkage of our filtered key genes with COVID-19. Of these, four key genes (CYBB, RPS3, RPS18, and RPS27) overlap with the COVID-19 pathway: CYBB is considered a major component of the phagocytic cells’ microbicidal oxidase system\cite{chan2013deficiency} and has been shown to produce excessive reactive oxygen species (ROS)\cite{violi2020sars} which are implicated in the exacerbation of clinical outcomes in severe COVID-19 infections\cite{elkahloun2020candesartan}. Overactivation of Nox2, associated with CYBB, has been linked to thrombotic complications in COVID-19 patients, suggesting a role for Nox2 as a mechanism favoring thrombotic-related ischemic events\cite{violi2020nox2}.

In addition, the Ribosomal Protein family has been reported as a potential inhibitor of COVID-19 replication\cite{rofeal2020ribosomal}, with RPS3, RPS18, and RPS27—members overlapping with our key network and the COVID-19 pathway—identified as critical. RPS3, in particular, was highlighted as a HUB gene in COVID-19\cite{prasad2021brain} and a potential therapeutic target\cite{chen2022mechanism}. COVID-19 exploits the mRNA-binding residues of RPS3 to influence both host and viral mRNA translation and stability\cite{havkin2023selective}.

Through the analysis of Gene Ontology\cite{gene2019gene} (Figure \ref{fig.6C}) and Uniport Keywords\cite{uniprot2023uniprot} (Figure \ref{fig.6D}), the majority of genes were enriched in immune-related keywords, further supporting the effectiveness and credibility of our method in selecting COVID-19 related genes.

The S100 Calcium Binding Protein Family was mentioned multiple times concerning immune-related keywords across databases and reported to play a role in the inflammatory conditions in COVID-19 patients with potential applications in predicting the severity of the disease\cite{bagheri2022prognostic,sattar2021s100,deguchi2021s100a8}. These genes have a closely related function with CYBB\cite{schenten2010ipla2}, potentially acting to activate CYBB  (the Activation of Phagocyte NADPH Oxidase, Nox2) in the immune response against COVID-19\cite{berthier2009important}.

Furthermore, in Uniport Keywords Ligand analysis, Calcium and Lectin emerge as key terms worth our attention. Literature suggests that COVID-19 critically impacts calcium balance and serves as a biomarker of clinical severity at the onset of symptoms\cite{zhou2020low,yang2021low}. Hypocalcemia has been strongly positively associated with COVID-19 severity\cite{alemzadeh2021effect}, and calcium channel blockers have been shown to reduce mortality from the disease\cite{crespi2021conflicts}. Lectin also demonstrates efficacy in combating and treating COVID-19\cite{ahmed2022plant,nascimento2021exploring,gupta2021status}.

These analyses indicate that the CIDER method effectively filters and identifies key genes with causal associations with COVID-19, unraveling intricate networks and pathways that offer a deeper and more comprehensive understanding of the disease's pathophysiology and contribute to identifying potential targets for therapeutic intervention.

\subsubsection{TCGA-LAML dataset}
We applied CIDER to the TCGA-LAML (acute myeloid leukemia) RNA-seq dataset\cite{anande2020rna}, which includes gene expression profiles from 179 individuals of three labels: Adverse, Intermediate and Favorable. For this analysis, we used only the samples labeled as Adverse and Favorable.

\subsubsection{Metric evaluation on the TCGA-LAML dataset}
Initially, we cleaned the gene expression data before using STRING and correction networks to construct the original gene network. Subsequently, we applied our CIDER methodology to identify and refine the top 1.5\% of highly variable genes from the original network. This approach resulted in the selection of the top 51 genes with the highest causal association scores, maintaining an 88.6\% accuracy compared to the original gene network. These key genes were then chosen for further biological analysis using tools such as the Database for Annotation, Visualization and Integrated Discovery (DAVID)\cite{huang2009david} and STRING database\cite{szklarczyk2023string}.

\begin{figure*}[ht]
    \centering
    \subfloat[STRING Network Subgraphs]{\includegraphics[width=.5\linewidth]{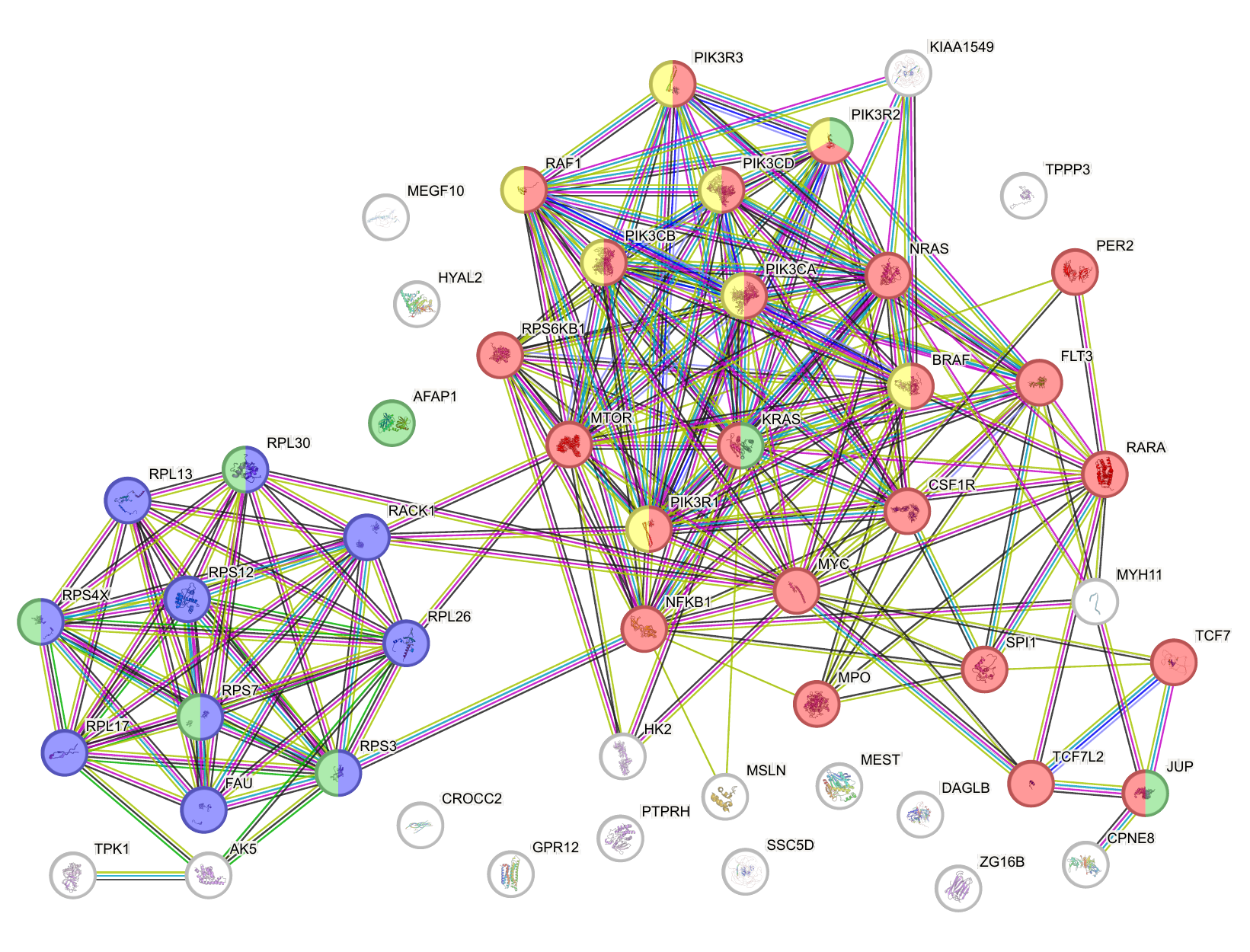} \label{fig.3A}}
    \subfloat[KEGG Pathway Enrichment]{\includegraphics[width=.5\linewidth]{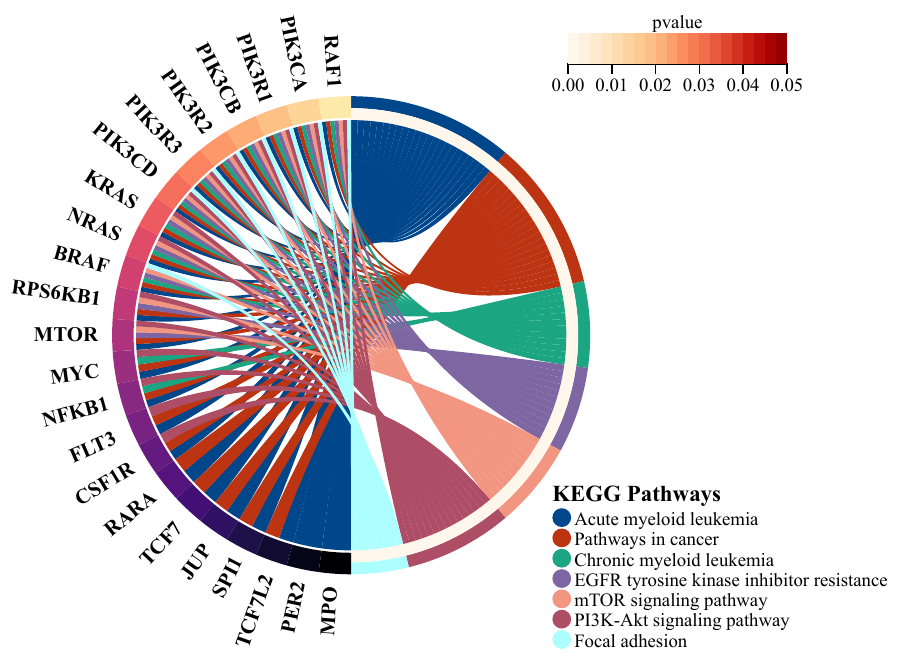}  \label{fig.3B}} 
    \hspace{2pt}
    \subfloat[GO Biological Process Enrichment]{\includegraphics[width=.5\linewidth]{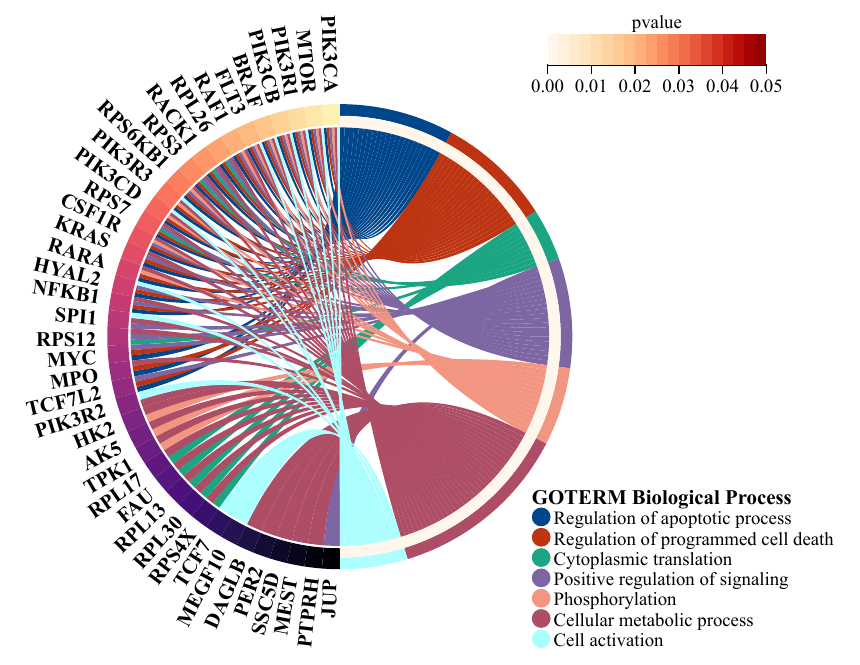}  \label{fig.3C}} 
    \subfloat[GO Cellular Component Enrichment]{\includegraphics[width=.5\linewidth]{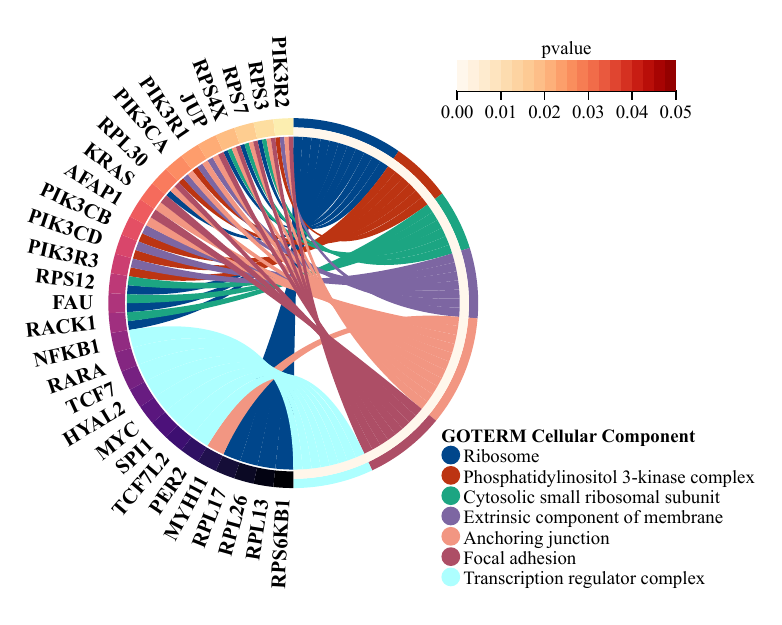}  \label{fig.3D}} 
    \hspace{2pt}

    \caption{TCGA-LAML analyses by CIDER. (a) Network subgraphs generated by STRING illustrate the division of 51 key genes into two distinct clusters, revealing potential unique functions and interactions within the molecular network related to acute myeloid leukemia (Red points) and ribosome (Blue points). (b) KEGG Pathways related to two types of myeloid leukemia, pathways in cancer, and cancer-related signaling pathways were enriched, suggesting a direct connection between the key genes and the leukemia mechanisms. (c) Gene ontology biological process enrichment analysis for the key genes identified by CIDER. (d) Gene ontology cellular component enrichment analysis for the key genes identified by CIDER.}
\end{figure*}

\subsubsection{Biological analysis for the TCGA-LAML dataset dataset}

For deeper insights into the key genes we found in TCGA-LAML with CIDER, we conducted enrichment analyses to find possible connections between the key genes and related terms (Figure \ref{fig.3B}-\ref{fig.3D}).

Among the key genes, the enrichment analysis from various databases indicated strong connections with leukemia. As shown in Figure \ref{fig.3B}, more than forty percent of the key genes are related to acute myeloid leukemia and the pathway in cancer\cite{kanehisa2000kegg}. Interestingly, 15 genes in the key genes are in terms of colorectal cancer, gastric cancer and breast cancer, and several research studies have indicated that patients undergoing chemotherapy and radiotherapy have an increased risk of developing acute leukemia\cite{rosner1978breast,valentini2011incidence,stein2012acute,wiggans1978probable,zhang2015secondary,konno1991advanced}.

In the network generated by the STRING database with a $score>0.4$ (Figure \ref{fig.3A}), there are two groups of genes contributing two significant subnetworks with dominant keywords Acute myeloid leukemia (KEGG Pathways, red points in Figure \ref{fig.3A}) and ribosome (Gene Ontology Cellular Component, blue points in Figure \ref{fig.3A} and Figure \ref{fig.3D}), indicating that the potential relationship between AML and ribosome\cite{marcel2017expression}.

Through the excavation of numerous research literature, we found that p53 pathway dysfunction is highly prevalent in acute myeloid leukemia\cite{quintas2017p53,zhang2017role,prokocimer2017dysfunctional}. Especially, RPL26, a ribosome gene in the left subnetwork (Blue points in Figure \ref{fig.3A}), is closely connected with MTOR and other AML-related genes. It plays a critical role in affecting the induction of p53\cite{takagi2005regulation,chen2012interactions}. RACK1, a gene associated with ribosomes, plays a significant role in promoting the proliferation of THP1 AML cells. This is achieved through the enhancement of glycogen synthase kinase 3$\beta$ (GSK3$\beta$) activity\cite{zhang2013rack1}. Additionally, overexpression of RACK1 is linked to chemotherapy resistance in T-cell acute lymphoblastic leukemia (T-ALL) cell lines. This resistance is mediated by increased protein kinase C$\alpha$ (PKC$\alpha$) activity, which helps protect the cells from chemotherapy-induced apoptosis\cite{lei2016increased}. These findings highlight the potential of targeting RACK1 in developing new therapeutic strategies for leukemia treatment.

The Ribosomal Protein S3 (RPS3) gene plays a significant role in the progression of acute lymphoblastic leukemia (ALL). Research has shown that over-expression of RPS3 promotes the growth and progression of all by down-regulating COX-2 through the NF-$\kappa$B pathway. This process involves suppressing COX-2 expression and its downstream targets, leading to increased cell proliferation and reduced apoptosis in leukemia cells. Targeting RPS3 could potentially be an effective therapeutic strategy for treating ALL\cite{hua2016over}.

The pathway and term of focal adhesion plays a significant role in the progression of AML\cite{carter2017focal,recher2004expression} (Figure \ref{fig.3D}). CIDER also found some important genes related to this term (Yellow and green points in Figure \ref{fig.3A} for KEGG focal adhesion pathway and GO Cellular Component focal adhesion, respectively). Targeting focal adhesion pathways with specific inhibitors has shown promise in preclinical studies, offering potential therapeutic strategies for AML treatment\cite{carter2015focal}.

In the larger subnetwork on the right side of the entire network (Figure \ref{fig.3A}), several genes not on the KEGG-AML pathway have also been reported to be associated with AML. KIAA1549L has indeed been implicated in the development of acute leukemia through its involvement in fusion genes. For instance, the PAX5-KIAA1549L fusion gene has been identified in pediatric B-cell precursor acute lymphoblastic leukemia (BCP-ALL), suggesting a role in leukemogenesis\cite{anderl2015pax5}. Additionally, KIAA1549L has been reported in another fusion gene, RUNX1-KIAA1549L, in a case of adult acute myeloid leukemia (AML), further highlighting its potential significance in leukemia development. These findings underscore the importance of KIAA1549L in the genetic landscape of acute leukemia\cite{abe2012novel}. The CPNE8 gene, part of the copine family, has been found to play a significant role in AML. Research by Ramsey and colleagues discovered that in AML patients, the CPNE8 gene can fuse with the AML1 gene, creating an AML-CPNE8 chimera. This fusion inhibits the transcription of AML genes, suggesting that CPNE8 may negatively regulate the proliferation of AML cancer cells\cite{ramsey2003fusion}.

The enrichment and biological analyses of key genes from TCGA-LAML data using CIDER have provided significant insights into the molecular underpinnings of acute myeloid leukemia (AML) and related leukemias.

\section{Discussion}

This work developed a new method "CIDER" of causal subgraph inference by implementing a Counterfactual-Invariant Diffusion process, i.e., with both counterfactual and diffusion implementations. In other words, it is able to generate causal explanations, which provides not only causal subgraphs due to counterfactual implementation but reliable causal links due to the diffusion process, which can be applied in various fields as a general method for interventional causal inference in contrast to the traditional association studies or observational causal inference from data. However, there are several remaining issues that need further improvement in this framework. 

\subsection{Unobserved confounders}
Handling unobserved confounders is a very important issue in causal inference, which limits the applications of many methods. In this work with the task model $\phi$, CIDER can actually reduce the influence of unobserved confounders. Let $E^u$ to denote the unobserved confounders, where $E^u$ must have some causal effect to $y$ and also to some part of $E^s$. We can split $E^s$ into $E^{su}$ and $E^{so}$ while $E^u$ has causal effect to $E^{su}$ but has no effect to $E^{so}$. Now we assume $E^c$ contributes more than $E^u$. According to the assumption of DAG, there are some other unobserved factors which make more influence on $E^{su}$, so it would make $do(E^c \cup E^{su})$ have less prediction ability for $y$ than $E^c \cup E^s$, which lets CIDER remove the $E^{su}$ from $E^c$, thus reducing the influence of unobserved confounders, and is also one major advantage of this work. 

\section{Conclusion}
This paper introduces CIDER for causal subgraph inference. Leveraging both counterfactual and diffusion implementations, CIDER provides generalizability in evaluating a subgraph's causal impact on labels. The counterfactual implementation ensures causal interpretation, while the diffusion process further enhances reliability. Real-world evaluations demonstrate the significant advantages of CIDER over existing methods: theoretically and computationally infer interventional causal subgraph (a set of causal edges) and alleviate unobserved confounder effects. Notably, CIDER's model-agnostic and task-agnostic nature offers broad applicability across various domains. Our work underscores the importance of explainability in fostering transparency and trust in black-box AI and aiding researchers in comprehending underlying phenomena and uncovering intricate network-level patterns. While CIDER theoretically possesses the capability to infer causal subgraphs in real-world scenarios even with unobserved confounders, further experiment is needed to fully evaluate this feature, highlighting a potential topic for future research. Additionally, investigating causal subgraph inference in hypergraphs presents an intriguing avenue for exploration and potential real-world application.

\section{Methods}
In this section, we introduce the new framework of CIDER. Different from previous association methods or observational causal works, CIDER is able to identify a causal subgraph for explaining both tasks 1) and 2) in the sense of counterfactual intervention. We first introduce an additive noise model and counterfactual invariant, and then a diffusion process to generate the causal subgraph by training the GNN based on the measured data ($X,y$). The overall framework of CIDER is shown in Figure \ref{fig.total}.


\subsection{Preliminaries and problem formulation}
As preliminaries, we define a few notations and concepts.
\subsubsection{Random graph}
A random graph is a useful mathematical tool for analyzing complex systems and networks in various fields \cite{mulet2002coloring, newman2003random, aittokallio2006graph, lesne2006complex, bringmann2019geometric}. In the machine learning field, random graphs have been used to model and analyze large-scale networks, such as social networks, biological networks, and the Internet.

As shown before, we study a random graph as $G = (X, V, E)$ where $|E|$ and $|V|$ are numbers of edges and nodes, respectively. Each node is a $d$-dimensional feature vector, and $X$ is a $|V| \times d$ matrix. One main purpose of this study is to investigate the conditional distribution of the edges, $P(E|X,V)$. We refer to this distribution as $P(E)$ for simplicity since we do not consider any other distribution of $E$.

\subsubsection{Graph neural networks and decision region} \label{sec.DecisionBound}
The overall prediction GNN Model\cite{kipf2016semi, velivckovic2017graph, xu2018powerful, mouli2021asymmetry} can be expressed as $\phi(G)=\phi_{fc}(\phi_{gc}(X, E))$, where $\phi_{gc}(X, E) \in \mathbb{R}^d$ represents the learned representations in the d-dimensional output space of the last convolutional layer. Here, $\phi_{gc}$ is the representation learning block that maps the input to the hidden space, and $\phi_{fc}$ is the prediction head that uses the learned representations for the specific task (e.g., classification or regression).

\cite{bajaj2021robust} models the decision region of the task GNN as a convex polytope in $\mathbb{R}^d$, encompassing numerous graph instances in the training set defined by a set of hyperplanes. This polytope aims to encompass graph instances with the same class label as many as possible.
 
Utilizing decision region theory, we can impose constraints to separate causal and spurious subgraphs, forcing counterfactual samples as close as possible to the original data, as illustrated in Figure \ref{fig.DecisionBound}. This approach helps to distinguish causal and spurious relations.

\subsection{Problem formulation}
Given a graph $G$ consisting of node features $X$, edges $E$, labels $y$, and a special task GNN $y=\phi(G)$, where $\phi$ is a task model. The goal is to find the subset $E^c$ that has the most causal effect on $y$. If $E$ is not available, we can use the measured $X$ to construct an association-based graph of nodes as an initial network instead. As mentioned above, there are two questions on the explainability \cite{amara2022graphframex}, i.e. 
1) causally explain the phenomena itself,  
2) transparently explain the given $\phi$. 
    For simplicity and without loss of generality, we drop $X$, so we can use $E$ to denote the graph $G$.

This paper focuses primarily on answering question 1) due to its great importance and inherent challenge in real-world applications. Note that in numerous scientific fields, including biology and physics, researchers are particularly interested in comprehending the "why" behind a phenomenon, namely its underlying causal subsystem or subgraph. 

We formulate the two explainabilities as a causal effect problem, with the difference being the task-GNN model condition. For addressing 1), we formulate Eq.(\ref{eq.asp1}) with an emphasis on the causal relationship between the causal subgraph $E^c$ of $E$ and the label $y$. We elucidate why $y$ is causally affected by $E^c$. The aim of this paper is to learn such a $\mathcal{F}^*$ in Eq.{\ref{eq.aim}}, which gives us the causal subgraph $E^c$.

\begin{eqnarray}
    \label{eq.aim}
    E^c &=& \mathcal{F}^*(E, y) \\ 
    \label{eq.asp1}
    P(y|E^c) &=& P(y|E)
\end{eqnarray}

To achieve 1) on causality rather than correlation, we must perform intervention or do-calculus on $E$. We use $E^s$ to represent the spurious subgraph of $E$, i.e., $E \setminus E^c$, in this paper. Hence, we formulate the problem using Eq.(\ref{eq.cfasp1}), so as to make the inference from association to causation. $CF(E^s)$ is defined as Eq.(\ref{eq.cf}), where $\tilde{E^s}$ is a counterfactual sample of $E^s$. Eq.(\ref{eq.cfasp1}) implies that we sample a new $E^s$ from the marginal distribution $P(E^s|E^c)$ without affecting the GNN's output, i.e. only changing $E^s$ without changing $E^c$ in this new pseudo-sample. This condition is stronger than doing an intervention on $E^c$. It is easy to prove that Eq.(\ref{eq.cfasp1}) is much stronger than Eq.(\ref{eq.asp1}) due to such a do-calculus or counterfactual implementation. If Eq.(\ref{eq.asp1}) is satisfied, $G^c$ is considered to only associate with $y$, rather than the cause of $y$, unless Eq.(\ref{eq.cfasp1}) is satisfied. 

\begin{eqnarray}
    \label{eq.cf}
    CF(E^s) &=& E^c \cup \tilde{E^s} \\
    \label{eq.cfasp1}
    P(y|CF(E^s)) &=& P(y|E) 
\end{eqnarray}

\subsection{Structural causal model and additive noise model}
An additive Noise Model (ANM) is a statistical model that describes the relationship between a signal and its noisy observations. ANMs are widely used in various machine learning fields, such as image processing and audio processing\cite{greenfeld2020robust, li2019certified}. The basic principle behind ANMs is that a signal of interest, $X$, is corrupted by random noise, $N$, resulting in the observed signal, $Y = \alpha X + N$. The presence of noise is essential in ANMs, as it enables us to identify which signal, $X$ or $Y$, is the cause and which is the effect. ANMs are used to understand the causal relationship between signals\cite{mooij2009regression}.

To identify the subgraph $E^c$ that has the most attribution to a given label $y$ in a graph $G$, it is necessary to assess the causal effect of $E^c$. However, this requires intervention in $E^s$. To tackle this challenge, we adopt an ANM to formulate the problem, as depicted by Eqs.(\ref{eq.zc}) and (\ref{eq.y}) where $E, y$ and $z$ represent edges, graph labels and hidden variables(graph representation) that generate $E$, respectively, and $z^c$ and $z^s|z^c$ are sampled from different normal distributions, and $f_E$ is VGAE decoder.
\begin{eqnarray}
    \label{eq.zc}
    E &=& f_E(z^c) \cup f_E(z^s) \\
    \label{eq.y}
    y &=& \phi(E^c) + \epsilon 
\end{eqnarray}

\subsection{Counterfactual invariant representation} \label{sec.CI}
Following \cite{mouli2021asymmetry}, the task-GNNs $\phi$ can be viewed to work as message-passing. We view $\phi$ through a message-passing lens. The aggregation process essentially acts as a sequence of transformers based on the edges in the feature matrix $X$. This perspective allows us to form the problem as an out-of-distribution (OOD) detection task. We treat each edge as a transformer applied to $X$, and aim to identify a set of edges that can be removed without making the resulting representation $\phi_{gc}(X)$ deviate from the original distribution. In other words, the $\phi_{gc}(X)$ should be a counterfactual invariant representation that only is sensitive to changes in the distribution of edges deemed causal($E^c$), not those deemed spurious($E^s$). This concept is illustrated in Figure {\ref{fig.DAG}, which depicts our method's directed acyclic graph (DAG). Note that $Z^c$ and $Z^s$ may exhibit correlations that are not considered here. Figure \ref{fig.DecisionBound}  further demonstrates how our approach integrates with the decision region framework discussed in Section \ref{sec.DecisionBound}.

\begin{figure}[!htbp]
    \centering
    \subfloat[DAG of our method]{\includegraphics[width=.4\linewidth]{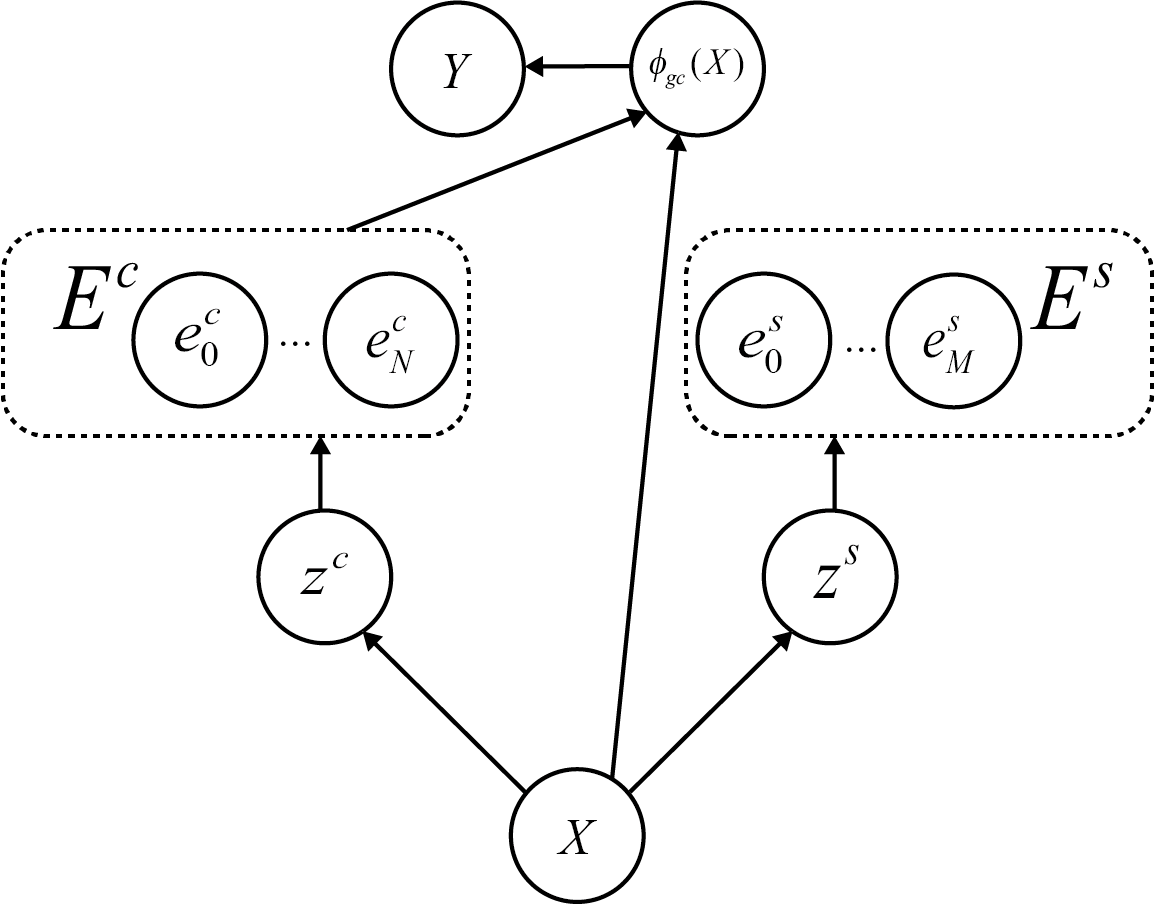} \label{fig.DAG}}  \hspace{2pt}
    \subfloat[Illustration of Decision Region]{\includegraphics[width=.55\linewidth]{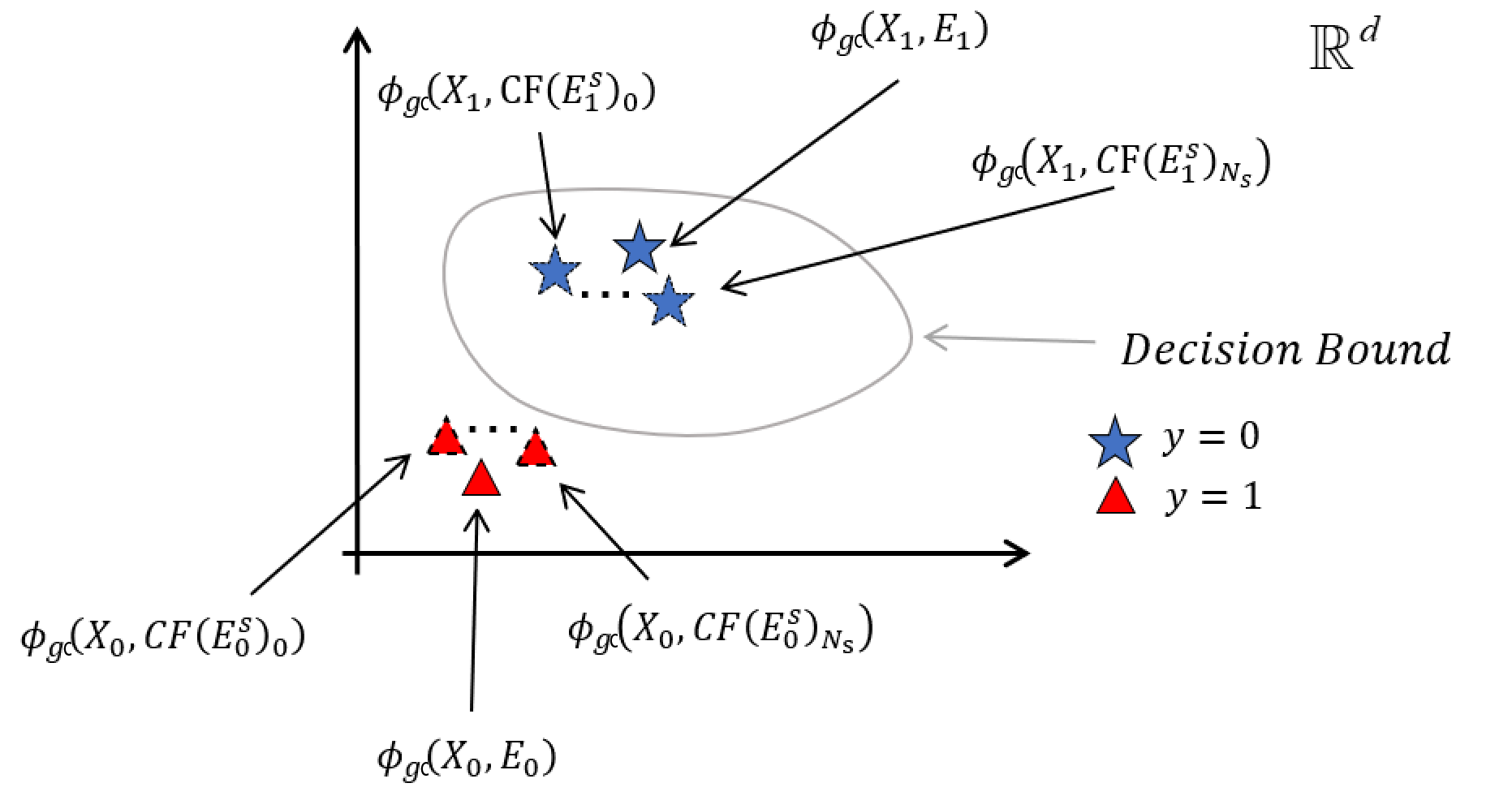}  \label{fig.DecisionBound}}
    \quad
    \subfloat[Illustration of the diffusion process for separating causal factors.]{
        \includegraphics[width=0.9\linewidth]{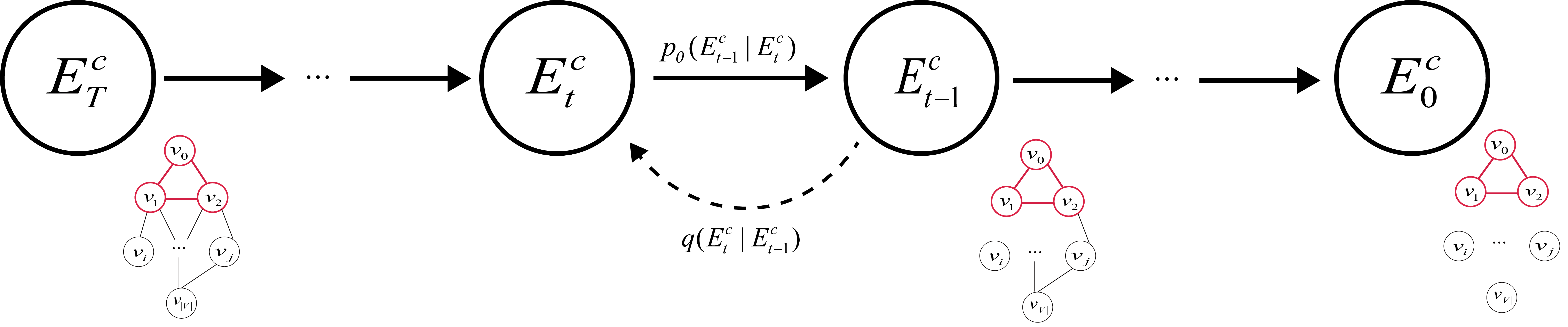}
        \label{fig.diffusion}
    }
    \caption{Causal structure of CIDER. (a) depicts the directed acyclic graph (DAG) of our method, where $E^c$ has a direct causal effect on $Y$, and $E^s$ represents spurious factors. (b) illustrates a simple 0-1 decision region of $E^c$ and $E^s$ in a 2D space, where the star and triangle indicate different graph labels, and their solid and dotted borders denote the original sample and the counterfactual sample in $E^s$, respectively. (c) is the illustration of the diffusion process for the separation of causal factors. The red subgraph has a direct causal effect on the given label.}
\end{figure}

\subsection{Distribution of causal subgraph}
In order to quantify the causal effect or strength of $E^c$, we need to perform an intervention. However, we cannot directly resolve the dependence between subsets of edges. Thus, we cannot use a mask on the original graph $G$ conducted in previous works \cite{ying2019gnnexplainer, bajaj2021robust, miao2022interpretable}, or calculate the causal effect of each edge separately \cite{lin2021generative}. Instead, we use the framework of random graphs to solve this problem at the distribution level. Although a graph may not look like other structured data, such as images or lists, it also has distributions. $P(E)$, $P(E^c)$, and $P(E^s)$ represent the distributions of $E$, $E^c$ (causal effect on $y$), and $E^{s}$ (no effect on $y|E^c$), respectively. The relationships between them can be formulated by Eq.(\ref{eq.2}). As mentioned in section \ref{sec.CI}, $E^c$ is a set of edges, which induces a series of transformations resulting in OOD of $\phi_{gc}(X)$ when their distribution changes. Therefore, we can use Variational Inference models to fit the distributions of $E^s$ and $E^c$, and obtain the subgraph of $E$ that contributes the most to the label $y$.
\begin{eqnarray} \label{eq.2}
    P(E) = P(E^c) P(E^{s}|E^c)
\end{eqnarray}

We can use a VGAE-like model to infer the distributions of $E^c$ and $E^s$, even if we do not have access to their actual distributions or samples. Using the counterfactual-invariant representation, we can obtain a partial distribution of $E^s$ without calculating the maximum likelihood of $E^c$ or $E^s$. Our goal is to ensure Eq.(\ref{eq.tv}) satisfied. By doing so, we can ensure that the set $E^s$ shown in Figure \ref{fig.DAG} has no causal effect on $y$ under the condition $E^c$.
\begin{eqnarray} \label{eq.tv}
    {\left | P(Y|\phi_{gc}(CF(E^s))) - P(Y|X) \right |}_{TV} = 0  
\end{eqnarray}


Although we can not obtain the total variation distance in Eq.(\ref{eq.tv}), we approximately minimize the L1 distance of the do-distribution (i.e., the distribution after do-calculus) and origin distribution on $\mathbb{R}^d$ to solve this problem. In other words, our goal is to minimize Eq.(\ref{eq.sample}). $N_c$ and $N_s$ denote the sampled numbers of the causal subgraph and spurious subgraph, respectively.
\begin{eqnarray} \label{eq.sample}
    L_1 = \sum_{i=1}^{N_c}\sum_{j=1}^{N_s} \left| \phi_{gc}(X, CF(E^s)) - \phi_{gc}(X, E) \right|
\end{eqnarray}

\subsection{Diffusion process}
A diffusion process is a mathematical model describing a quantity's spread over a specified area. Diffusion processes are used to model a range of phenomena, including the dispersion of pollutants in the environment \cite{ukpaka2016empirical}, the transmission of infectious diseases \cite{setti2020potential}, and the propagation of information in social networks \cite{bakshy2012role}. Diffusion models have recently emerged as powerful generative models in various machine learning applications \cite{ho2020denoising,sohl2015deep,song2019generative}. The mathematical theory of diffusion processes is based on the principles of probability theory and stochastic processes. The diffusion process can be described as a random walk in which a particle moves randomly in each step. These rules determine the probability of the particle making a particular move and the time taken for the particle to move.

Due to the complexity of $E^s$ and the lack of a real causal subgraph, we cannot infer the distribution at one step. In order to obtain a reliable causal subgraph, we make our framework work as a diffusion process. We assume that the origin set $E$ is equal to the final causal subgraph $E^c_0$ plus $T$ steps' noise, so let $E^c_T$ to denote the initial set $E$, $E^c_{T-1}$ and $E^s_{T-1}|E^c_{T-1}$ to denote the result of the first separation of $E^c_T$, where the $E_c^{T-1}$ contains some non-causal edges, implying that the distribution $P(E^c_{T-1})$ is actually equal to $P(E^c_{T-2}, E^s_{T-2})$. Hence, we do the same inference process as before, and then we can obtain the $E^c_{T-2}$ and $E^s_{T-2}|E^c_{T-2}$ distribution, but $E^c_{T-2}$ also includes non-causal factors. Therefore, we repeat the process recurrently. Letting $E^c_{T-i}$ and $E^s_{T-i}|E^c_{T-i}$ to denote the results of the $i$-th separation process, we can write such a recurrent process as Eq.(\ref{eq.recurrent}).
\begin{eqnarray} \label{eq.recurrent}
    P(E^c_T)&=&P(E^c_{T-1}) P(E^s_{T-1}|E^c_{T-1}) \\ \nonumber
    &=&P(E^c_{T-2}) P(E^s_{T-2}|E^c_{T-2}) P(E^s_{T-1}|E^c_{T-1}) \\ \nonumber
    &=& \cdots \\ \nonumber
    &=&P(E^c_0)\prod_{t=0}^{T-1} P(E^s_t|E^c_t)
\end{eqnarray}

In order to simultaneously infer $P(E^c_t)$ and $P(E^s_t|E^c_t)$, we utilize a two-channel VGAE as shown in Figure \ref{fig.total}. We use Eq.(\ref{eq.loss}) as the loss function to train the CIDER. $L_1$, mentioned in Eq.(\ref{eq.sample}), is a loss to calculate the causal effect of $E^c$ on $y$. $L_{kld}$ is the KL-divergence of $z^c$ and $z^s$ with the normal distribution. $L_{reconstruction}$ is the reconstruction loss of $E^c \cup E^s$ and $E$ in the form of cross-entropy in our model. We use the $L_{task}$  to avoid $E^s$ always equal to $\varnothing$ and $E^c$ always equal to $E$.

\begin{eqnarray} \label{eq.loss}
    L_{CIDER} &=& L_1 + L_{VGAE}  \\
    L_{VGAE} &=& L_{kld} + L_{reconstuction} \\
    \label{eq.task}
    L_{task} &=& mse(\phi(E^c, X), y)
\end{eqnarray}

\subsubsection{Diffusion process in hidden space}
The graph's distribution of edges differs from that of the normal diffusion process because the edge is not a typical structured data, making it difficult to add and remove noise from it. However, as shown in Eq.(\ref{eq.zdiffusion}), the hidden variable $z$ is observed as a diffusion process, which resembles a normal diffusion process. $\theta$ denotes the parameters of CIDER. In fact, our model treats $z$ as a diffusion process, similar to DALLE2\cite{ramesh2022hierarchical}, and actually, the edge reconstruction makes the diffusion model trainable, although the noise distribution is unknown.

\begin{eqnarray} \label{eq.zdiffusion}
	P(Z^c_t|Z^c_{t+1})&=&\mathcal{N}(\mu^c_\theta(E^c_t), \Sigma^c_{\theta}(E^c_t)) \\
	P(Z^c_0|Z^c_T)&=&P(Z^c_0|E^c_1) P(E^c_1|Z^c_T) \\ \nonumber
	&=&P(Z^c_0|Z^c_1) P(Z^c_1|Z^c_T) \\ \nonumber
	&=&P(Z^c_0|Z^c_1) P(Z^c_1|Z^c_2) P(Z^c_2|Z^c_T) \\ \nonumber
	&=&\cdots \\ \nonumber
	&=&\prod_{t=1}^{T-1} P(Z^c_{t-1}|Z^c_{t})
\end{eqnarray}

The pseudocode for CIDER is shown in Algorithm \ref{alg.cider}. At each step, we sample $z_t^c$ and $z_t^s$ from the distributions $P(z_t^c|z_{t+1}^c)$ and $P(z_t^s|z_{t+1}^s)$, respectively. Then, we use $z_t^c$ and $z_t^s$ to update $E_{t-1}^c$ and $E_{t-1}^s$. Subsequently, we use $E_{t-1}^c$ and $E_{t-1}^s$ to update $z_{t-1}^c$ and $z_{t-1}^s$. This aforementioned process is repeated until $t=0$.

\begin{algorithm}[H]
	\caption{CIDER with recurrent causal distillation}
	\label{alg.cider}
	\begin{algorithmic}
		\State {\bfseries Input:} data ${(X, E)}$
		\For{$t=T$ {\bfseries to} $1$}
		\State $z^c_t \thicksim \mathcal{N}(\mu^c_\theta(E^c_t), \Sigma^c_{\theta}(E^c_t))$
		\State $z^s_t|z^c_t \thicksim \mathcal{N}(\mu^s_\theta(E^c_t), \Sigma^s_{\theta}(E^c_t))$
		\State $E^c_{t-1} = sigmod((z^c_t)(z^c_t)^T$)
		\State $E^s_{t-1} = sigmod((z^s_t|z^c_t)(z^s_t|z^c_t)^T$)
        \State $\theta = \theta - \nabla L(E^c_{t-1}, E^s_{t-1}, y, X)$
		\EndFor
	\end{algorithmic}
\end{algorithm}

\subsection{Analysis process on biological data}
Figure \ref{fig.application} outlines the workflow of CIDER for inferring the causal genes and cell types associated with a disease state from scRNA-seq data. Initially, we perform data cleaning on the original gene-cell expression matrix, denoted as $X_{\text{origin}}\in \mathbb{R}^{\# \text{genes} \times \# \text{cells}}$. The data cleaning involves removing cells or genes that exhibit zero expression values, resulting in a selected gene-cell matrix, $X_{\text{selected}}\in \mathbb{R}^{\# \text{selected genes} \times \# \text{selected cells}}$. Subsequently, we construct a gene network, $E_{\text{gene}}$, and a cell type network, $E_{\text{cell type}}$. 

\begin{figure}[ht]
  \centering
  \includegraphics[width=1\linewidth]{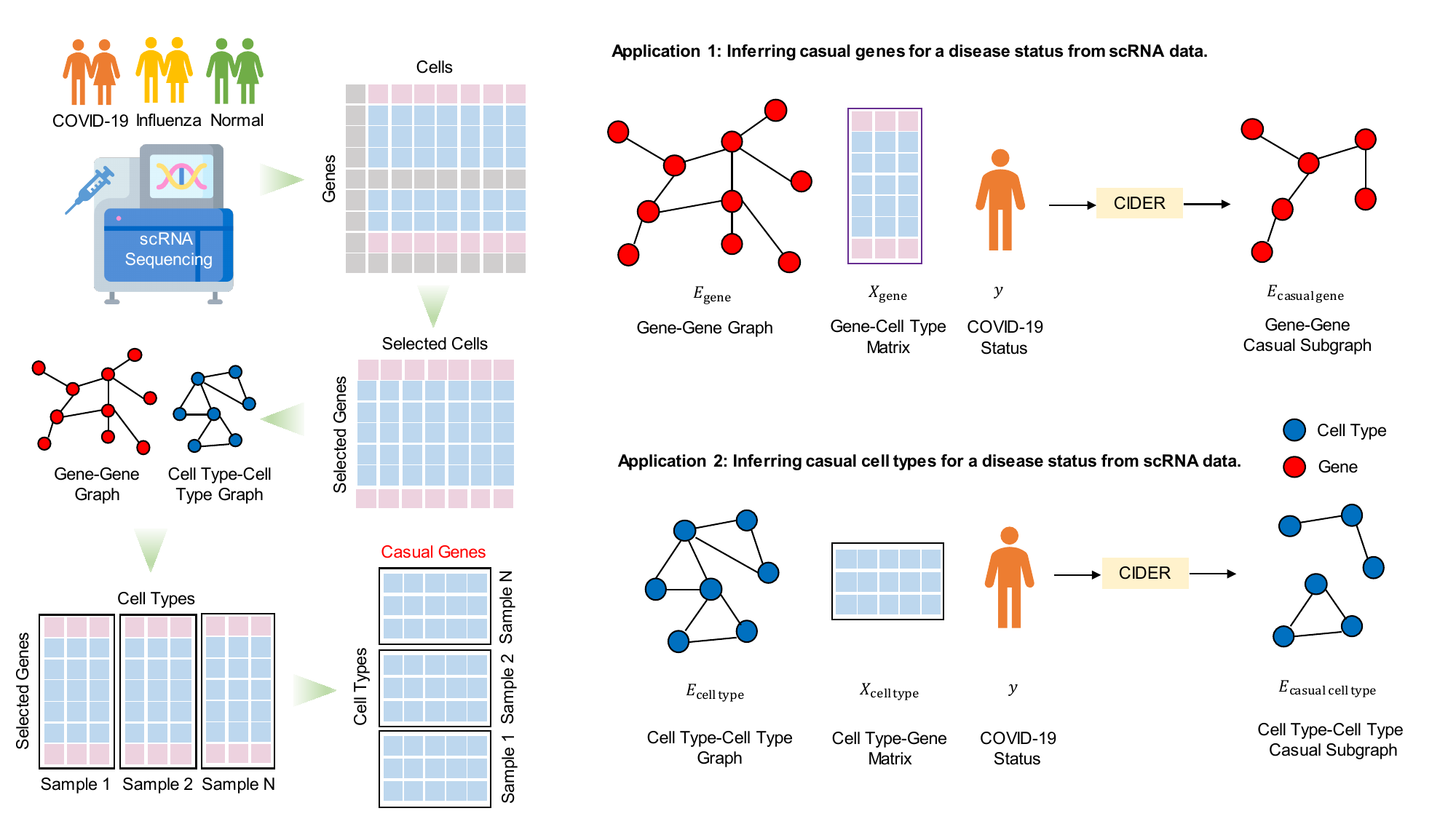}
  \caption{Workflow of CIDER for inferring the causal genes and cell types associated with a disease state from scRNA-seq data}
  \label{fig.application}
\end{figure}

To infer the causal genes for each blood sample, using $X_{\text{selected}}$, we computed a gene-cell type matrix $X_{\text{gene}}\in \mathbb{R}^{\# \text{selected genes} \times \# \text{cell types}}$, where a row represents a selected gene, and a column represents a cell type. The matrix entry is the average gene expression of a selected gene across all cells in a given cell type. We consider the disease state of a blood sample as the task label $y$. By leveraging $E_{\text{gene}}$, $X_{\text{gene}}$, and the labels $y$, CIDER can be employed to generate explanation $E^c_{\text{gene}}$, indicating the causal effects of genes for COVID-19 and influenza.

To infer the causal cell types for each blood sample, we construct a cell type-gene matrix $X_{\text{cell type}}\in \mathbb{R}^{\# \text{cell type} \times \# \text{casual genes}}$, where a row represents a causal gene, and a column represents a cell type. The matrix entry is the average gene expression of a given causal gene across all cells in a given cell type. We fit $E_{\text{cell type}}$, $X_{\text{cell type}}$, and the label $y$ into CIDER, allowing us to determine a causal cell type network $E^c_{\text{cell type}}$. This network facilitates the analysis of not only the impact of cell types on COVID-19 but also the combined effects of various cell types. 

\backmatter

\bibliography{sn-bibliography}
\newpage

\begin{appendix}
\section{Further Implemetation Details}
\textbf{Datasets statistics} MUTAG and NCI1 are used to test CIDER's performance. The statistics of these datasets are shown in Table \ref{tab:dataset}.
\begin{table}[htbp]
\centering
\caption{Statistics of datasets.}
\label{tab:dataset}
\begin{tabular}{c||ccc}
\toprule
Dataset & \#Graphs & \#Nodes & \#Edges \\ \midrule
MUTAG & 4,337 & 30.32 & 30.77 \\
NCI1 & 4,110 & 29.87 & 32.30 \\
\bottomrule
\end{tabular}
\end{table}

\textbf{Exprienmt Environment} all experiments could be conducted on a server with only one NVIDIA RTX 3080ti GPU, 2.2GHz Intel Xeon Gold 6148 CPU and 256GB RAM.

\textbf{Experimental Details} For the training of CIDER, we use Adam optimizer with learning rate 0.001 and weight decay 0.0005. The batch size is set to 128. The number of training epochs is 500. The number of diffusion steps is 10. The number of Causal Sample $N_c$ is 1, the number of Spurious Sample $N_s$ is 4.

\textbf{Baseline Models} GNNexplainer and PGExplainer use the pyg version provided by pyg. GEM and OrphicX use the office version provided by author.

\end{appendix}

\end{document}